\documentclass[final,5p,times,twocolumn]{elsarticle}

\usepackage{amssymb}

\journal{Astroparticle Physics}

\begin{document}

\begin{frontmatter}



\title{A Simple Parallelization Scheme for Extensive Air Shower Simulations}


\author[label1]{B.~T.~Stokes\corref{cor1}}
\ead{stokes@cosmic.utah.edu}

\author[label1]{R.~Cady}

\author[label1,label2]{D.~Ivanov}

\author[label1]{J.~N.~Matthews}

\author[label1]{G.~B.~Thomson}

\address[label1]{University of Utah Department of Physics \& Astronomy and High Energy Astrophysics Institute, Salt Lake City, Utah 84112, USA}
\address[label2]{Department of Physics and Astronomy, Rutgers---The State University of New Jersey\\ Piscataway, New Jersey 08854, USA}
\cortext[cor1]{Corresponding author}

\begin{abstract}
A simple method for the parallelization of extensive air shower simulations is 
described.  A shower is simulated at fixed steps in altitude.  At each step,
daughter particles below a specified energy threshold are siphoned off and 
tabulated for further simulation.  Once the entire shower has been tabulated, 
the resulting list of particles is concatenated and divided into
separate list files where each possesses a similar projected computation time.  
These lists are then placed on a 
computation cluster where the simulation can be completed in a piecemeal fashion 
as computing resources become available.  Once the simulation is complete, the
outputs are reassembled as a complete air shower simulation.  The original 
simulation program (in this case CORSIKA) is in \emph{no way altered} for this 
procedure.  Verification is obtained by comparisons of $10^{16.5}$~eV showers 
produced with and without parallelization. 

\end{abstract}


\end{frontmatter}


\section{Introduction}
\label{sect:intro}

In the past 50 years, much progress has been made in the understanding of 
Extensive Air Showers (EAS) associated with Ultra-High Energy Cosmic Rays 
(UHECRs).  However, the historical
difference in energy determination between Surface Detection 
(SD)~\cite{volcano:spectrum}\cite{haverah:spectrum}\cite{yakutsk:spectrum}\cite{agasa:spectrum} and 
Fluorescence Detection 
(FD)~\cite{flyseye:spectrum}\cite{hiresmono:gzk}\cite{hiresstero:spectrum}
has yet to be resolved.  In its hybrid mode, the Pierre Auger 
experiment~\cite{auger} reports a 30\% discrepancy for 
simulation based energy determination between SD and FD 
for events observed in hybrid operation mode~\cite{auger:fdsd}.  

We posit that this 
discrepancy could be better understood if it were not for the fact that 
it has been computationally infeasible to simulate large numbers of 
EAS with primary energy 
$>10^{18}$~eV without utilizing statistical
thinning methods that fully simulate only a small representative fraction of 
the EAS.  These thinned simulations certainly can be adequate for 
calculating longitudinal profiles~\cite{Pryke:2000ac} and average lateral 
distributions~\cite{livni:1}.  However, they 
neither capture the full breadth of fluctuations at the distance scale of 
individual surface detector counters nor do they provide all of the specific 
particle information necessary to properly estimate counter energy deposition
and the consequent electronic response.

Non-thinned simulations of UHECRs are very computationally intensive.
Using a single modern CPU core, 
the simulation of the EAS for a single $10^{17}$~eV 
proton requires on the order of $1$~day. 
Simulation times increase more or less linearly 
with energy.  Extrapolated to the logical extreme, this implies that, 
without continued progress in computational 
ability, the simulation of the largest UHECR observation reported thus far 
($3.2\times10^{20}$~eV~\cite{flyseye:spectrum}) could span the better part of
a decade.

One solution for this computational deficiency is parallelization.
By dividing the task between many different CPU cores, a simulation that 
previously would have taken years to conclude can be completed in days or even 
hours.  By employing scoring strategies to optimize the division of labor, it
is possible to simulate even the largest observed 
EAS with nothing more than the 
spare computational power that inadvertently arises in the scheduling of large
jobs on computational clusters. 

This paper is the first of three to describe methods used in the Telescope
Array (TA) Collaboration~\cite{ta:1,ta:2} to simulate EAS as seen by the TA 
surface detector (TASD).  
The second paper will deal with ``dethinning,'' that is,
replacing the shower particles eliminated in the thinning process.  The third
paper will describe the simulation of the TASD response to EAS and show
comparisons between the actual TASD data and a spectral set of simulated EAS.

\section{Parallelization Overview}
\label{sect:parallel}

Simulation programs for EAS are particularly well-suited for 
parallelization because they do not contain self-interaction.  That is, 
the individual daughter particles in the shower interact exclusively with 
atmospheric medium and not with each other.  Thus, each EAS can be thought of as
the superposition of many smaller sub-showers.  Parallelization can then be
carried out in the following steps:
\begin{enumerate}
\item Initially, a single computer is utilized to 
separate the EAS simulation into many smaller, more manageable, simulations 
by running the simulation repeatedly through small steps 
in atmospheric depth.  
\item At each step, the simulation output is sorted with 
particles above a nominal upper threshold being passed back through the 
simulation. Particles below a lower variable threshold are discarded  
and the rest of the output is appended to a master list.  
\item Eventually, all of the simulated particles fall below the 
nominal upper threshold. The master list contains all the input
parameter sets necessary for a series of simulations that can be superimposed 
to reconstitute the the original EAS.  
\item The master list is then divided into 
sub-lists and divided amongst a larger number 
of computers either manually or via 
clustering.  
\item When all of the sub-list simulations are finished, 
the final total simulation can then be reassembled.
\end{enumerate}

A critical aspect of this procedure is that the actual simulation source code 
is in 
no way altered.  All aspects of parallelization are achieved by translating 
each generation of simulation output files into the next generation of 
simulation input files via a series of scripts and compiled programs under the 
direction of a master script which explicitly tracks spatial and temporal 
information for each component simulation.

\section{Parallelization Application}
\label{sect:corsika}

While this is, in principle, a fairly straightforward procedure, it is important
to consider how this procedure works in practice for a particular simulation
package.  For this purpose, we use CORSIKA v6.960~\cite{corsika}.  High energy hadronic interactions are modeled by QGSJET-II-03~\cite{qgsjet}, low energy hadronic interactions are modeled by FLUKA2008.3c~\cite{fluka1,fluka2}, and 
electromagnetic interactions are modeled by EGS4~\cite{egs4}. 

As a first step, a CORSIKA ``input card'' is created for the primary particle. 
While this input card can encompass a wide range of CORSIKA options, there are 
some important constraints:
\begin{enumerate}
\item There is only one EAS in the simulation run. 
\item The starting point of the arrival time scale is set to the first 
interaction.  This greatly simplifies tracking time offsets later in the 
procedure
\item The zenith angle of the primary particle does not exceed $60^\circ$.
This constraint is necessary in order to keep the primary zenith angles of the
secondary EAS below $70^\circ$.  Above $70^\circ$, CORSIKA uses a curved
earth model which greatly complicates the propagation of temporal and 
spatial offsets through the parallelization.
\item The observation level is set high in the atmosphere (e.g. 80~km).  By 
doing so, we minimize the simulation time for the linear stage of the 
procedure. 
\item The ground level core location is set to have the same x-y coordinate 
system as the EAS.
\end{enumerate}

Once an input card is created, it is submitted to the CORSIKA simulation package.
The particles in the simulated output are then each assigned a score,
$T_i$ which corresponds to the maximum time necessary to simulate a 
complete EAS down to the final observation level for that particle. The value of $T_i$ was determined using the following relation: 
\begin{equation}
T_i\propto E_i\times e^{t_{si}/\tau},
\end{equation} 
where $E_i$ is the energy of the particle, $t_{si}$ is the atmosphere slant depth between the position of the particle and the ground, and $\tau=50$~g/cm$^2$.
While this is a crude estimate of simulation time at best, it is fairly accurate insofar as the EAS maximum does not occur too far above the ground.

The particles are then divided into two categories: $T_{i} > T_{max}$ and 
$T_{i} < T_{max}$, where $T_{max}$ is the maximum time to completion for 
individual jobs in parallel portion of this process.  Particles where 
$T_{i} < T_{max}$ are added to a master list.  Each line of the list contains:
$T_i$, particle type, energy, trajectory, position, time, and random number 
seeds.

For particles where $T_{i} > T_{max}$, the time and position 
are noted in a separate
file and a new CORSIKA input card is generated.
The new card has an observation level 50~g/cm$^2$ lower in slant depth than the 
starting altitude.  The resulting set of input cards is then submitted to the 
CORSIKA package for simulation.  The resulting particles have their times and
positions offset by the initial values recorded above.  This process undergoes 
many iterations until there are no particles where $T_{i} > T_{max}$.

Once the shower is decomposed into particles where $T_{i} < T_{max}$, a cut is 
applied on particles where $E_{i} < t_{vi}*(1$~MeV$\cdot$cm$^2$/g) with $t_{vi}$
being the vertical depth between the particle and the ground level.  This 
eliminates particles which are sufficiently low in energy that subsequent 
propagation would not
be expected to persist to ground level due to either 
ionization and/or further EAS production.  The 
resulting list of particles is then sorted into sub-lists where 
$\sum_iT_i\simeq T_{max}$.

The simulation can now be carried out in a parallel fashion.  For each sub-list, CORSIKA input cards are created  
for each individual particle.  This differs from the non-parallel portion 
above in that every EAS is simulated all the way to ground level.  Once 
the EAS are simulated for every particle, 
the time and position offsets for each particle is applied to the 
respective CORSIKA output file and then all the EAS from the sub-list are
concatenated into a single CORSIKA output file.  
From there, the final step is to further concatenate the 
concatenated sub-list outputs 
from the parallel jobs into a single CORSIKA output file.   

The predominate advantage of this procedure lies in its inherent flexibility.
Because each sub-list is independent in its execution, the simulation can 
proceed on whatever number of computational nodes is available and even on 
multiple systems.  Because the 
size of each sub-list can be controlled by simply increasing or decreasing 
$T_{max}$ it is very simple to make use of whatever excess capacity might be  
available due to scheduling gaps in a large computational cluster.  
Furthermore, it would be trivial to adapt this method to volunteer computing.

\section{Parallelization Validation}
\label{sect:ver}

\begin{figure*}[t,b,p]
\begin{center}
(a)\includegraphics[width=0.96\textwidth]{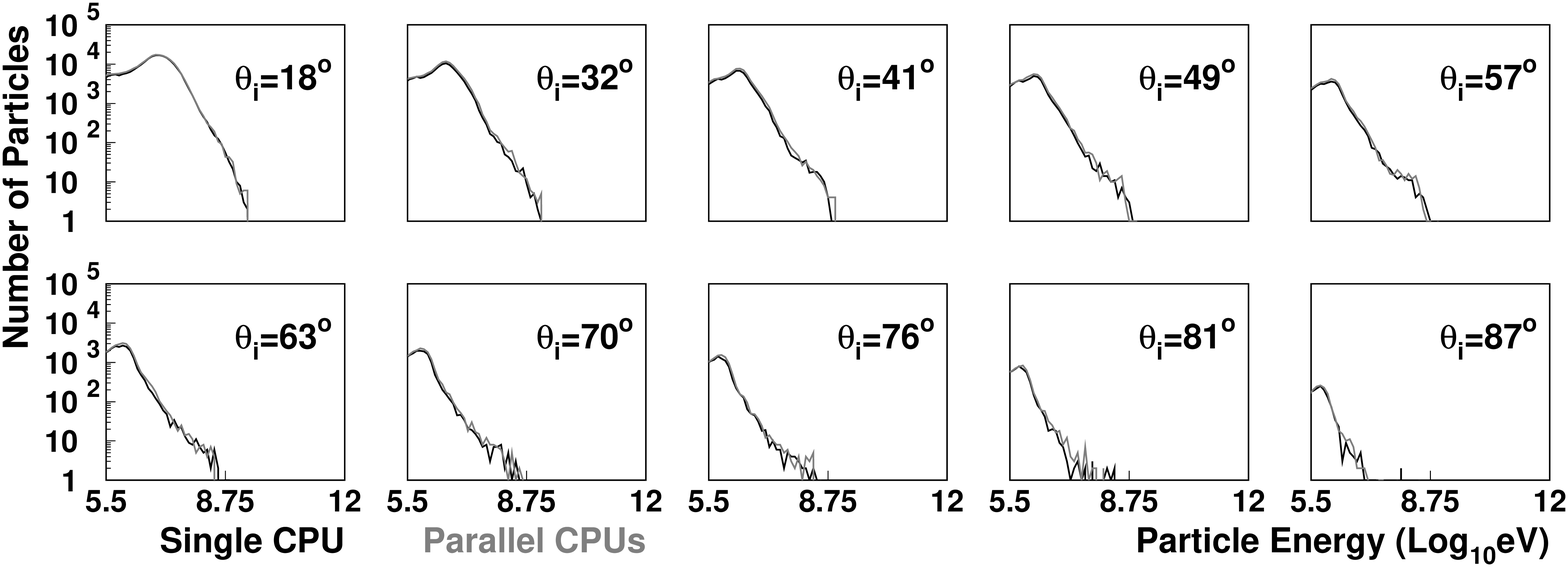}\\
(b)\includegraphics[width=0.96\textwidth]{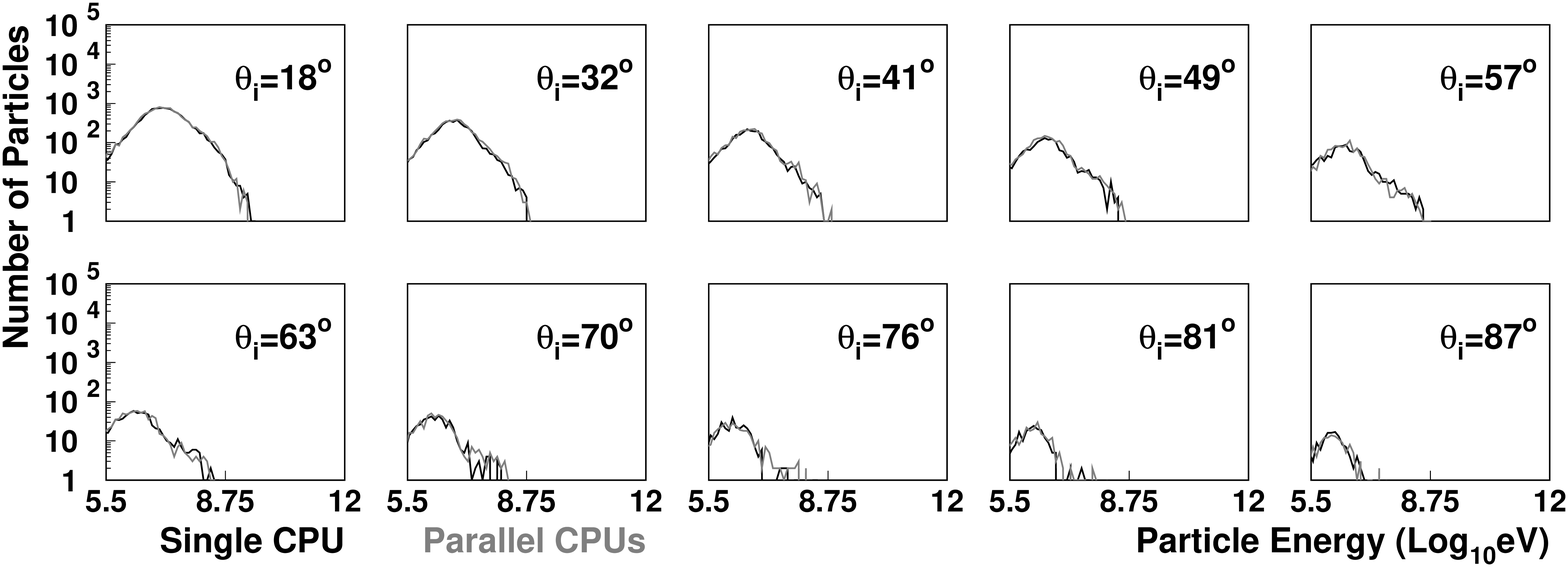}\\
(c)\includegraphics[width=0.96\textwidth]{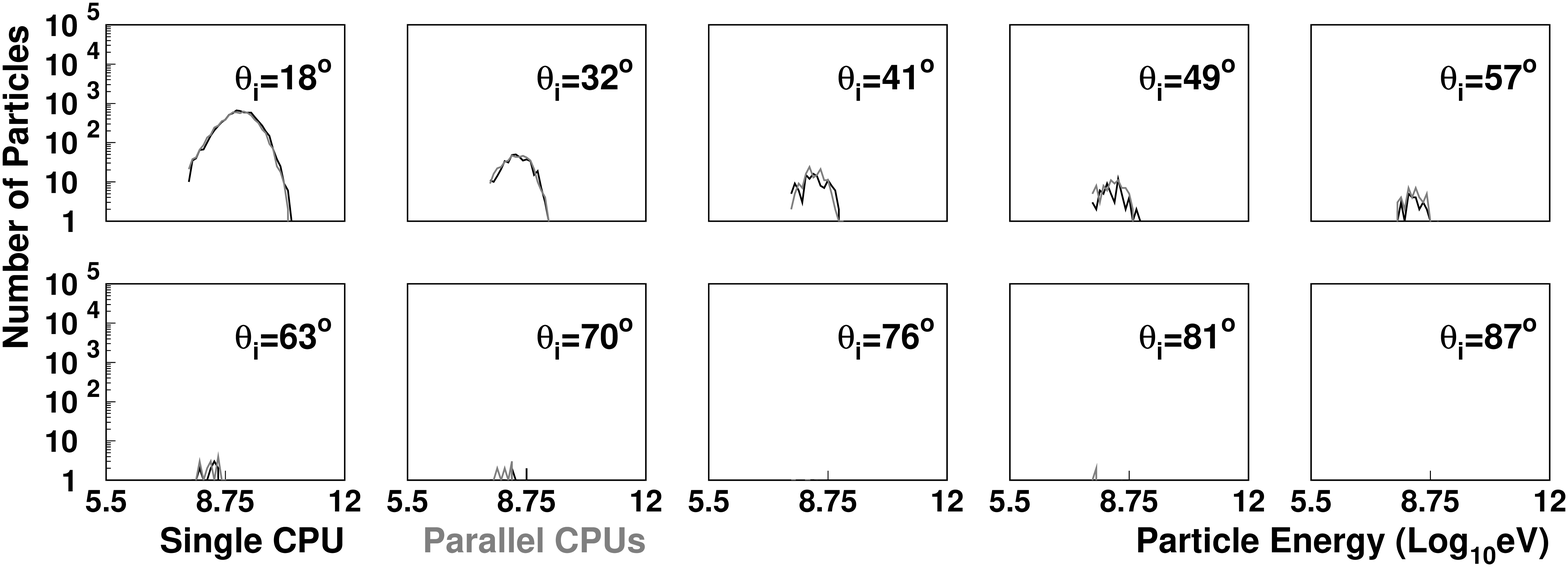}
\end{center}
\caption{Comparison of secondary particle spectra for two protonic showers 
(one generated with a single CPU and one generated in parallel with many CPUs) 
with primary energy {\boldmath$E_0=10^{16.5}$~{\bf eV} and primary zenith angle 
$\theta_0=0^\circ$}. For each shower, 
simulated particles whose ground position was within
a region enclosed by shower rotation angles $\Phi=[-30^\circ,30^\circ]$ and 
lateral distances $r=[500{\rm m},1000{\rm m}]$ were tabulated with respect
to particle type, incident angle with respect to the ground, $\theta_i$, and 
kinetic energy.  The resulting spectra are shown in $\cos\theta_i=0.1$ increment
bins for a) photons, b) electrons and positrons, and c) muons.  For each 
histogram, good agreement is observed between simulations generated linearly
(black) and via parallelization (gray).}
\label{fig1}
\end{figure*}

\begin{figure*}[t,b,p]
\begin{center}
(a)\includegraphics[width=0.96\textwidth]{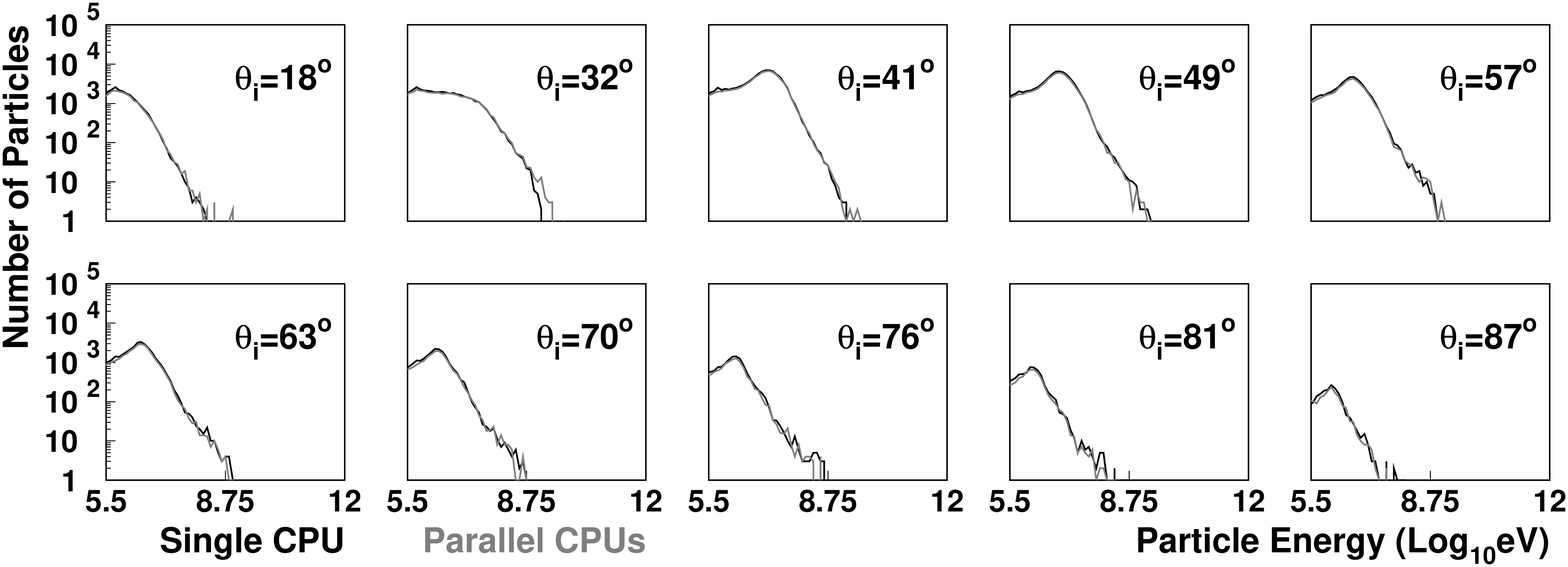}\\
(b)\includegraphics[width=0.96\textwidth]{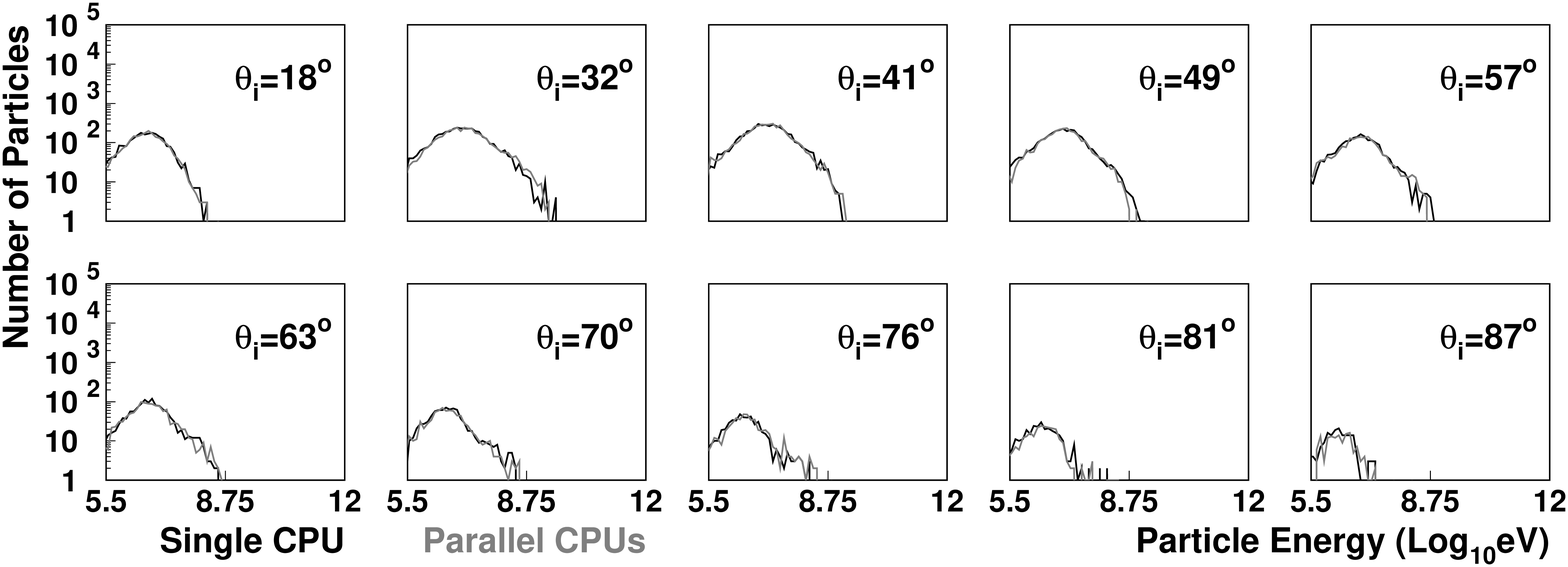}\\
(c)\includegraphics[width=0.96\textwidth]{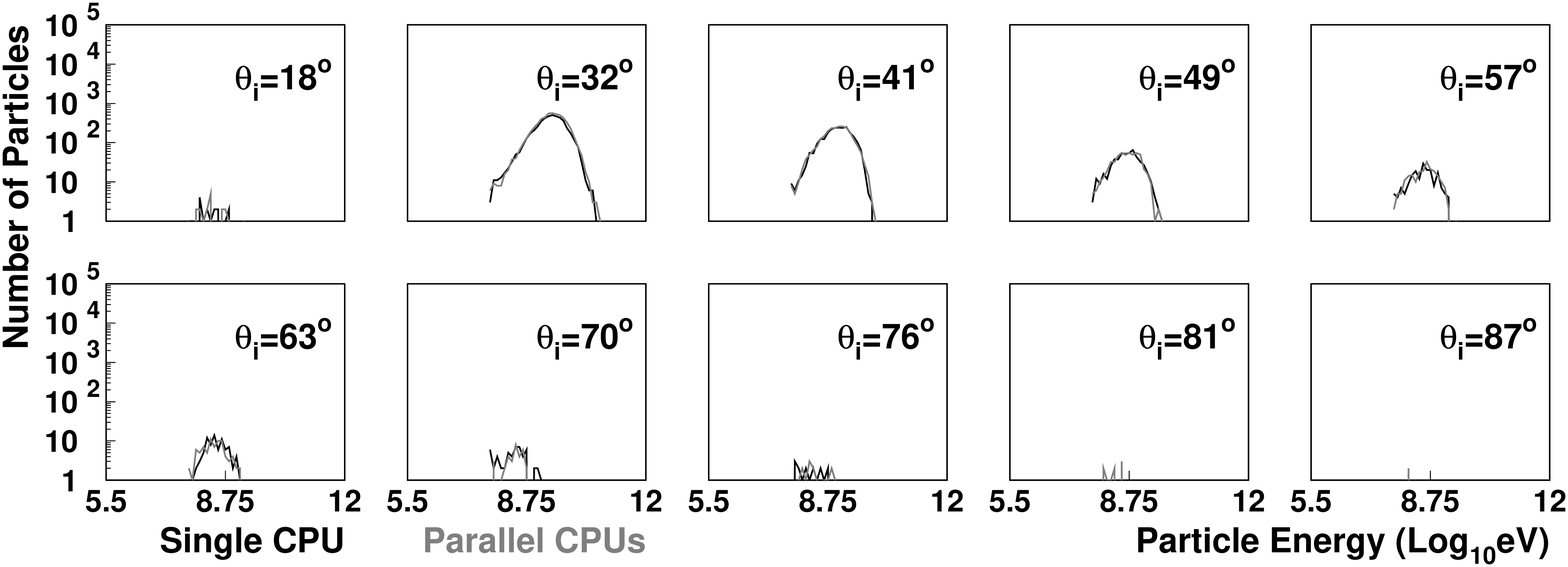}
\end{center}
\caption{Comparison of secondary particle spectra for two protonic showers 
(one generated with a single CPU and one generated in parallel with many CPUs) 
with primary energy {\boldmath$E_0=10^{16.5}$~{\bf eV} and primary zenith angle 
$\theta_0=30^\circ$}. For each shower, 
simulated particles whose ground position was within
a region enclosed by shower rotation angles $\Phi=[-30^\circ,30^\circ]$ and 
lateral distances $r=[500{\rm m},1000{\rm m}]$ were tabulated with respect
to particle type, incident angle with respect to the ground, $\theta_i$, and 
kinetic energy.  The resulting spectra are shown in $\cos\theta_i=0.1$ increment
bins for a) photons, b) electrons and positrons, and c) muons.  For each 
histogram, good agreement is observed between simulations generated linearly
(black) and via parallelization (gray).}
\label{fig2}
\end{figure*}

\begin{figure*}[t,b,p]
\begin{center}
(a)\includegraphics[width=0.96\textwidth]{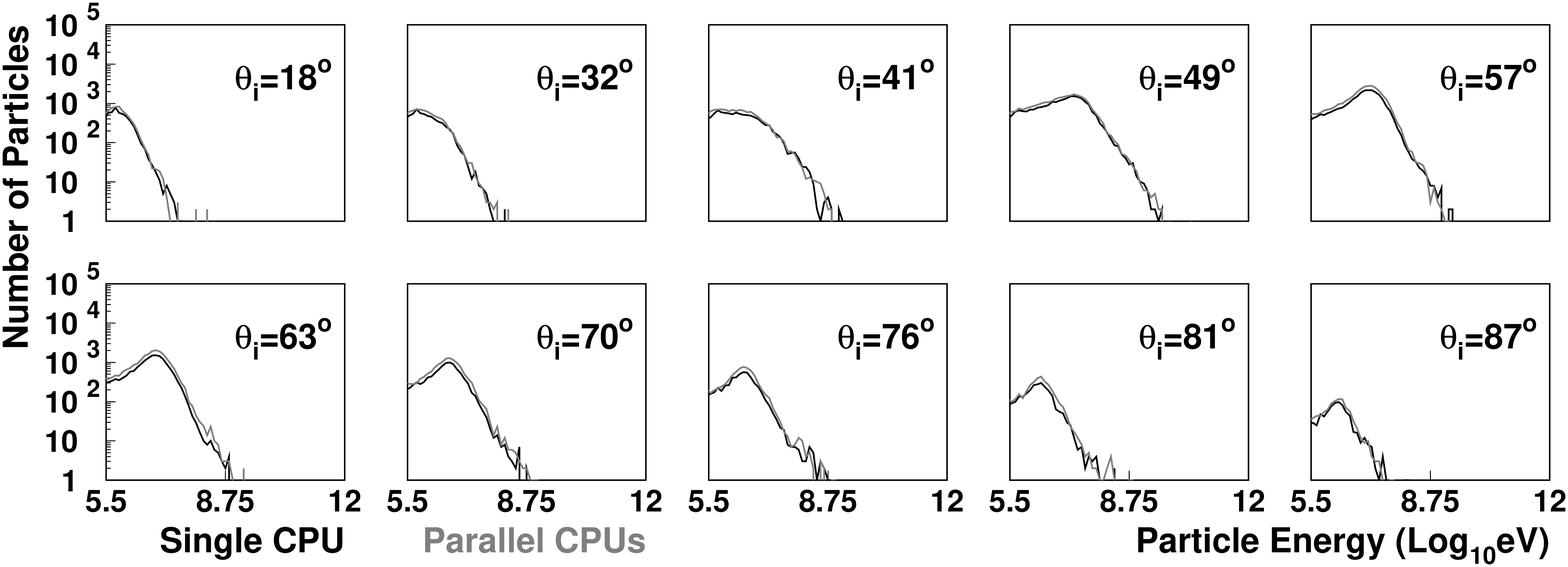}\\
(b)\includegraphics[width=0.96\textwidth]{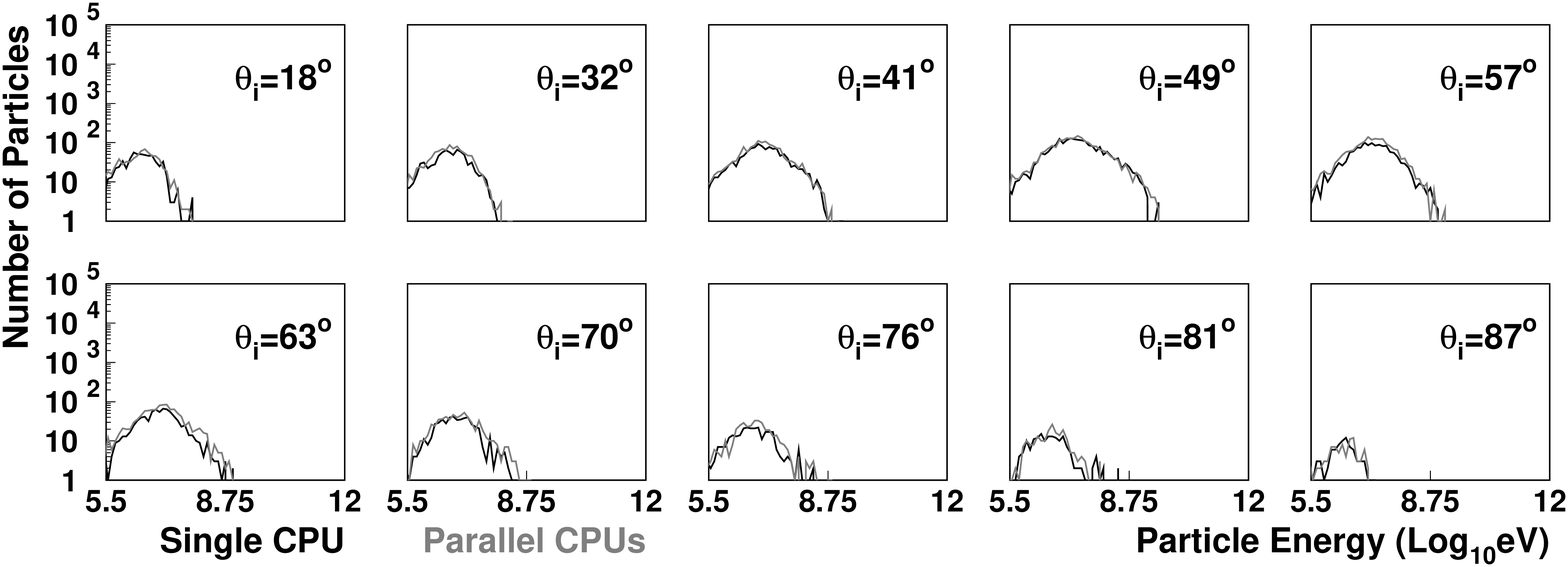}\\
(c)\includegraphics[width=0.96\textwidth]{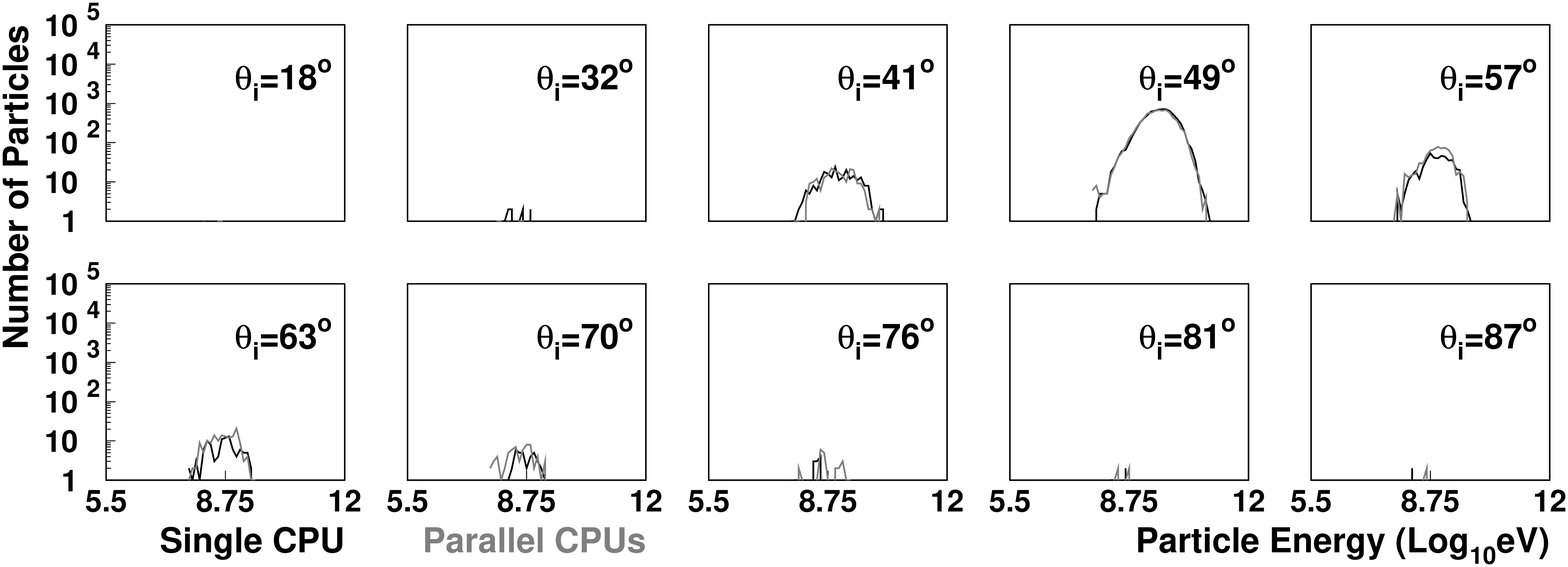}
\end{center}
\caption{Comparison of secondary particle spectra for two protonic showers 
(one generated with a single CPU and one generated in parallel with many CPUs) 
with primary energy {\boldmath$E_0=10^{16.5}$~{\bf eV} and primary zenith angle 
$\theta_0=45^\circ$}. For each shower, 
simulated particles whose ground position was within
a region enclosed by shower rotation angles $\Phi=[-30^\circ,30^\circ]$ and 
lateral distances $r=[500{\rm m},1000{\rm m}]$ were tabulated with respect
to particle type, incident angle with respect to the ground, $\theta_i$, and 
kinetic energy.  The resulting spectra are shown in $\cos\theta_i=0.1$ increment
bins for a) photons, b) electrons and positrons, and c) muons.  For each 
histogram, good agreement is observed between simulations generated linearly
(black) and via parallelization (gray).}
\label{fig3}
\end{figure*}

\begin{figure*}[t,b,p]
\begin{center}
(a)\includegraphics[width=0.96\textwidth]{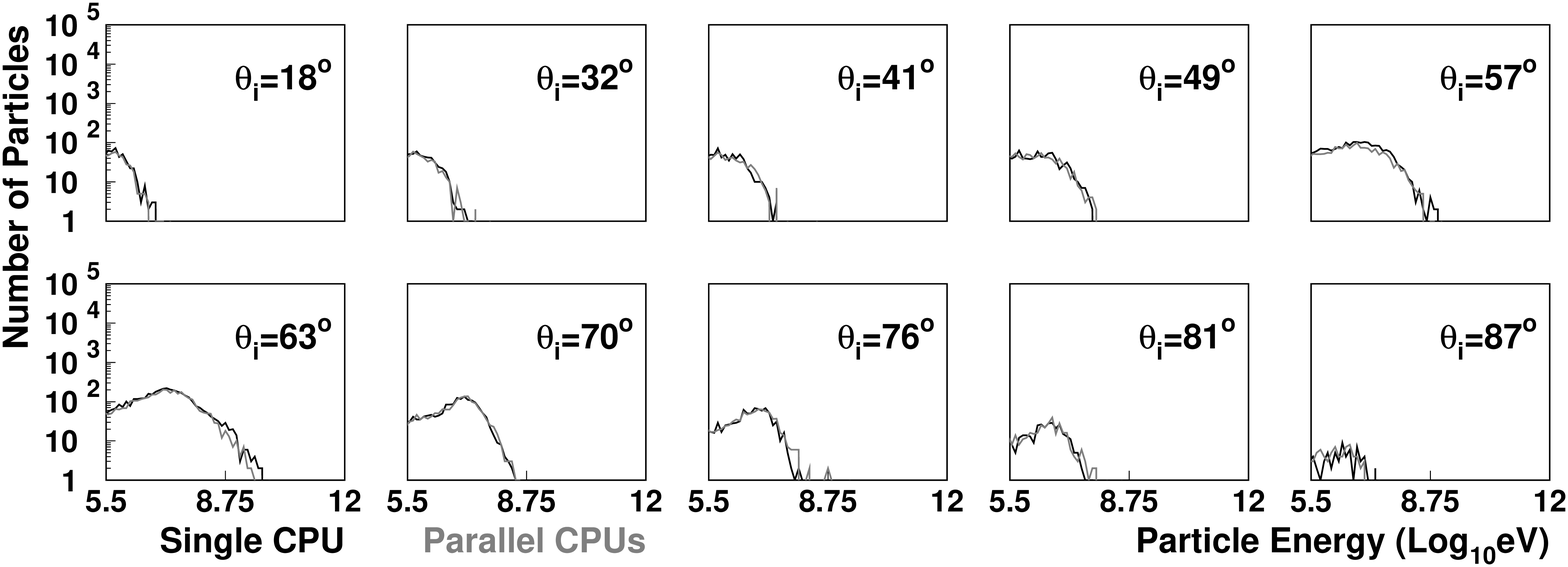}\\
(b)\includegraphics[width=0.96\textwidth]{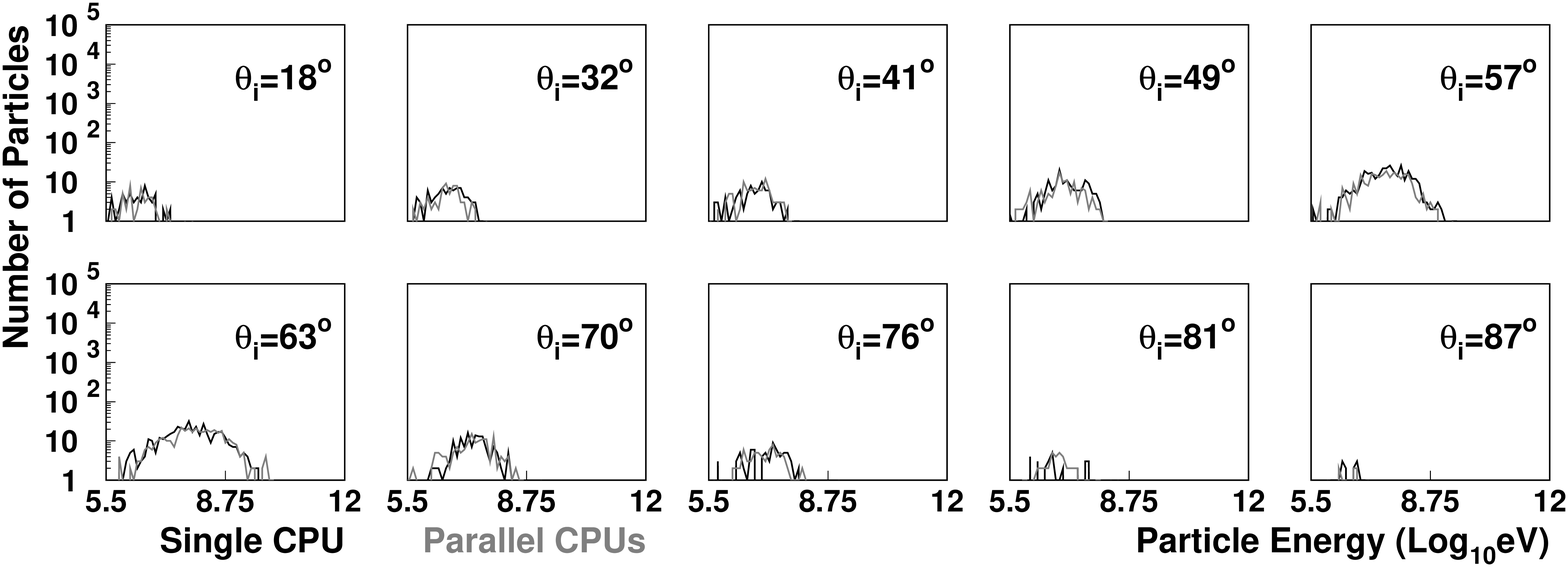}\\
(c)\includegraphics[width=0.96\textwidth]{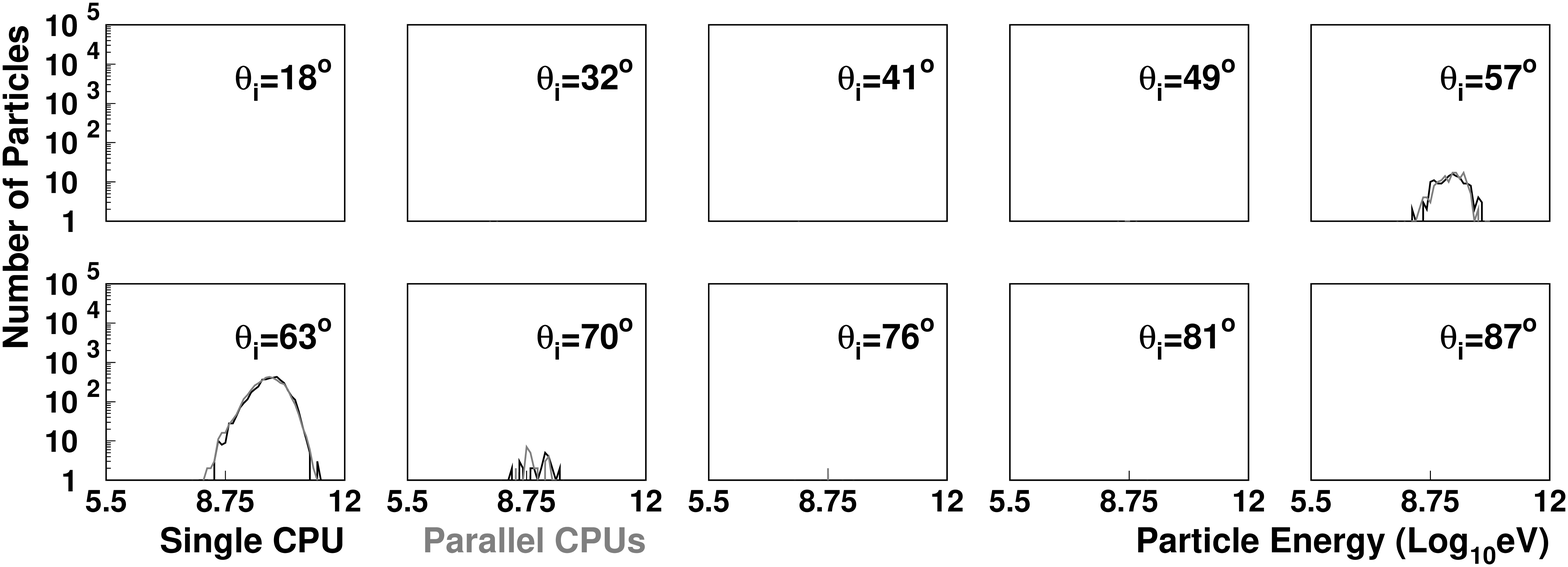}
\end{center}
\caption{Comparison of secondary particle spectra for two protonic EAS 
simulations 
(one generated with a single CPU and one generated in parallel with many CPUs) 
with primary energy {\boldmath$E_0=10^{16.5}$~{\bf eV} and primary zenith angle 
$\theta_0=60^\circ$}. For each shower, 
simulated particles whose ground position was within
a region enclosed by shower rotation angles $\Phi=[-30^\circ,30^\circ]$ and 
lateral distances $r=[500{\rm m},1000{\rm m}]$ were tabulated with respect
to particle type, incident angle with respect to the ground, $\theta_i$, and 
kinetic energy.  The resulting spectra are shown in $\cos\theta_i=0.1$ increment
bins for a) photons, b) electrons and positrons, and c) muons.  For each 
histogram, good agreement is observed between simulations generated linearly
(black) and via parallelization (gray).}
\label{fig4}
\end{figure*}

The parallelization method is validated by comparing pairs of EAS simulated 
with the
same input parameters both with and without parallelization.  For comparison
purposes, we 
initiated with a $10^{16.5}$~eV proton at a fixed height of first interaction of
30~km. This comparatively low primary energy was chosen due to the 
time required to generate EAS without parallelization.  Four different 
primary zenith angles were selected:  $0^\circ$, $30^\circ$, $45^\circ$, and $60^\circ$. In Figures~\ref{fig1}-\ref{fig4},
we show comparisons of secondary spectra from CORSIKA EAS 
simulated with and without parallelization. While the same random number seeds
were used for the single-process simulations and the top level of the
parallelized simulations, this can only guarantee that the interactions
down to the first observation level of the parallelization scheme (in this 
case 29~km) will be the same for the two simulations.  Taking this built-in
discrepancy into account, the simulations with and without parallelization 
agree remarkably well for Figures~\ref{fig1}-\ref{fig4}. 

\begin{figure*}[t,b,p]
\begin{center}
(a)\includegraphics[width=0.96\textwidth]{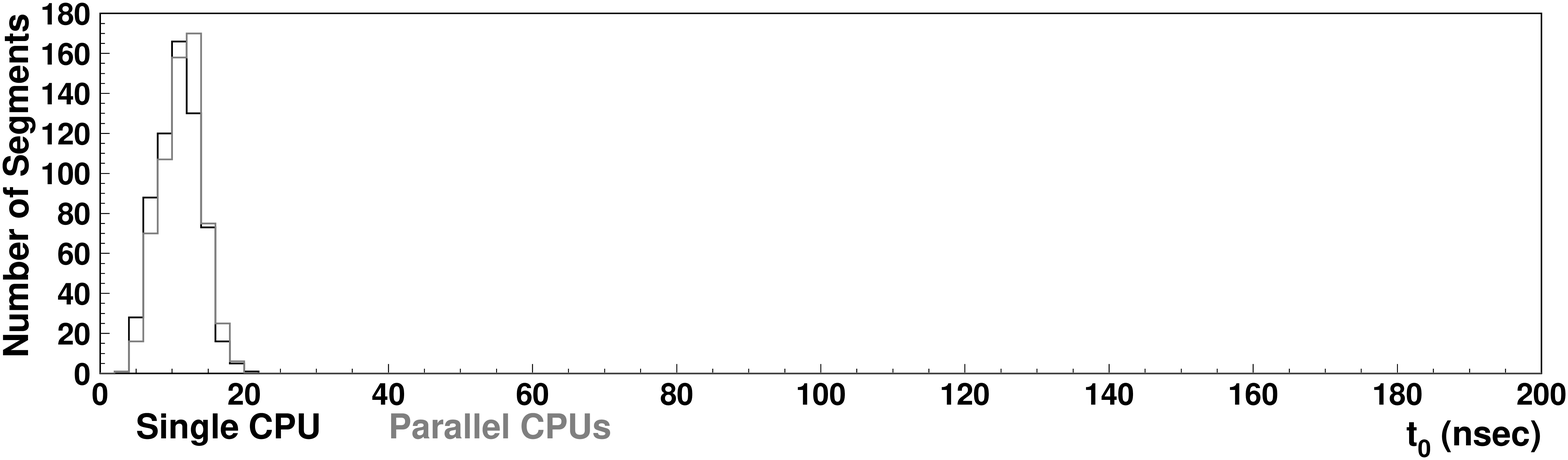}\\
(b)\includegraphics[width=0.96\textwidth]{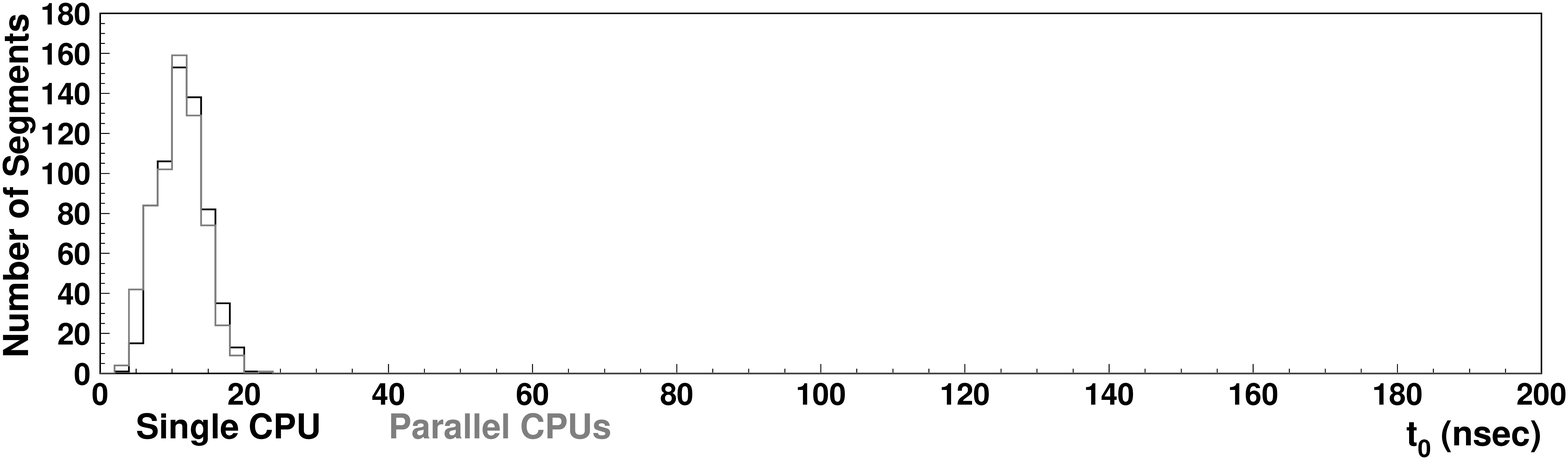}\\
(c)\includegraphics[width=0.96\textwidth]{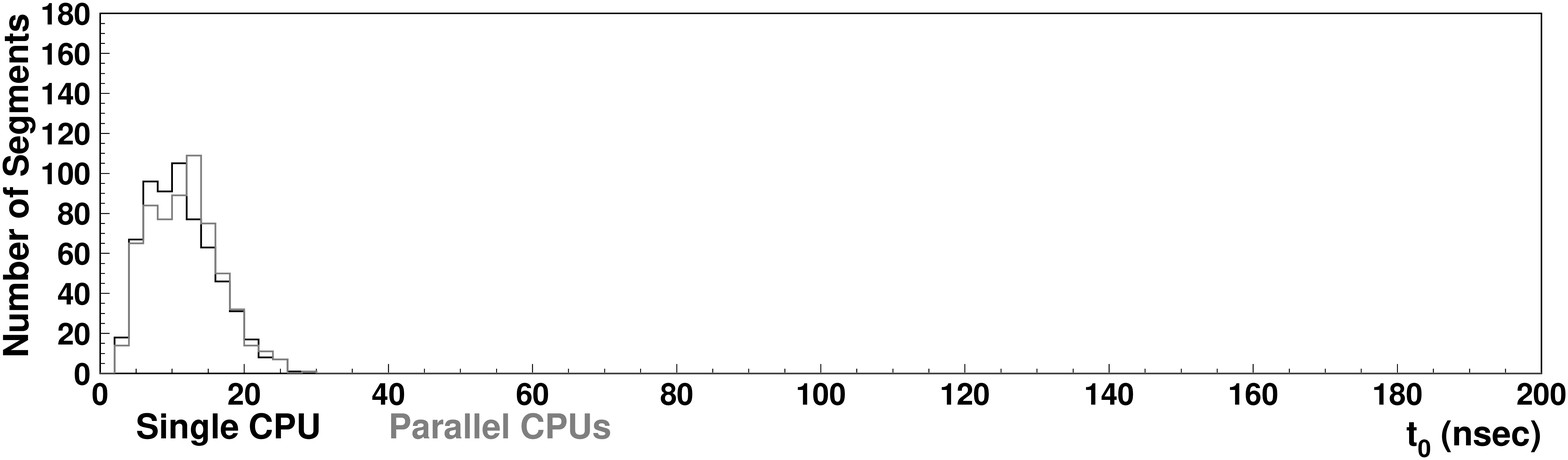}\\
(d)\includegraphics[width=0.96\textwidth]{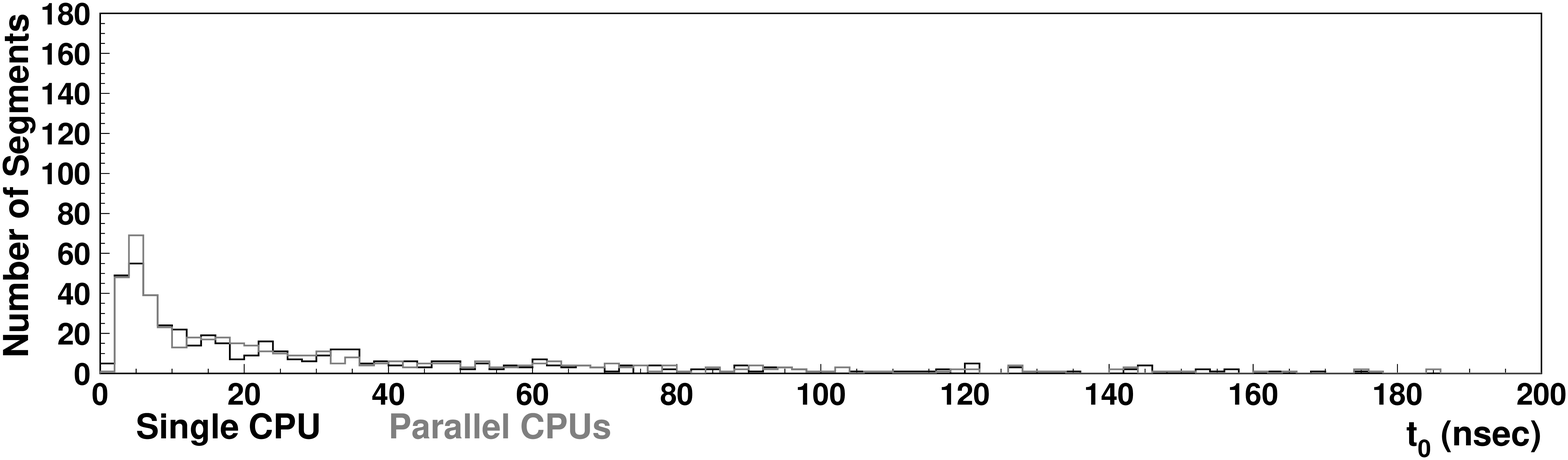}
\end{center}
\caption{Comparison of distribution of initial arrival times, $t_0$, for $2\times 2$~m segments in plane normal to shower trajectory for $10^{16.5}$~eV 
protonic EAS 
simulations with primary zenith angles of a) $0^\circ$, b) $30^\circ$, 
c) $45^\circ$, and d) $60^\circ$.  In each case, $t_0$ was measured for segments
200~m lateral distance from the shower core.   For each 
histogram, good agreement in both mean value and variance are observed 
between simulations generated linearly
(black) and via parallelization (gray).}
\label{fig5}
\end{figure*}

\begin{figure*}[t,b,p]
\begin{center}
(a)\includegraphics[width=0.96\textwidth]{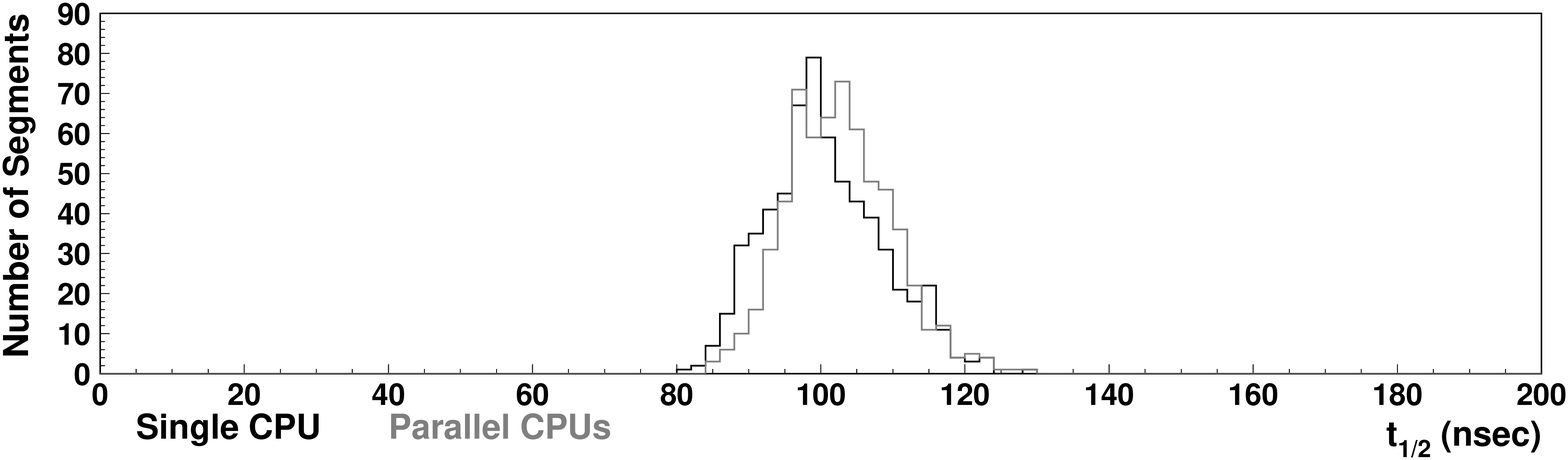}\\
(b)\includegraphics[width=0.96\textwidth]{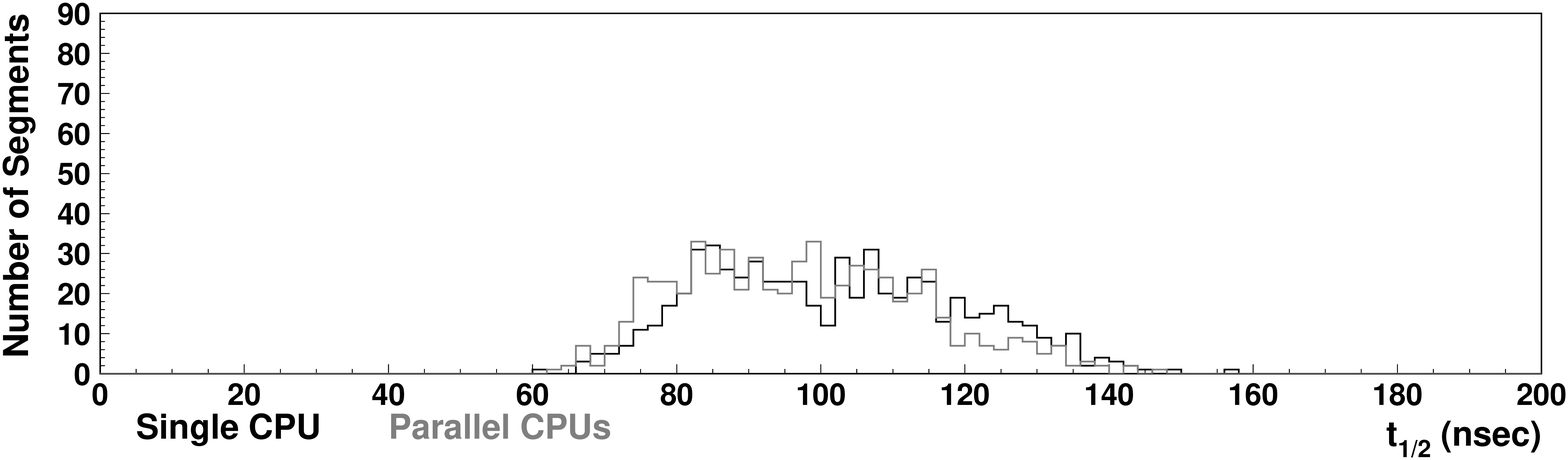}\\
(c)\includegraphics[width=0.96\textwidth]{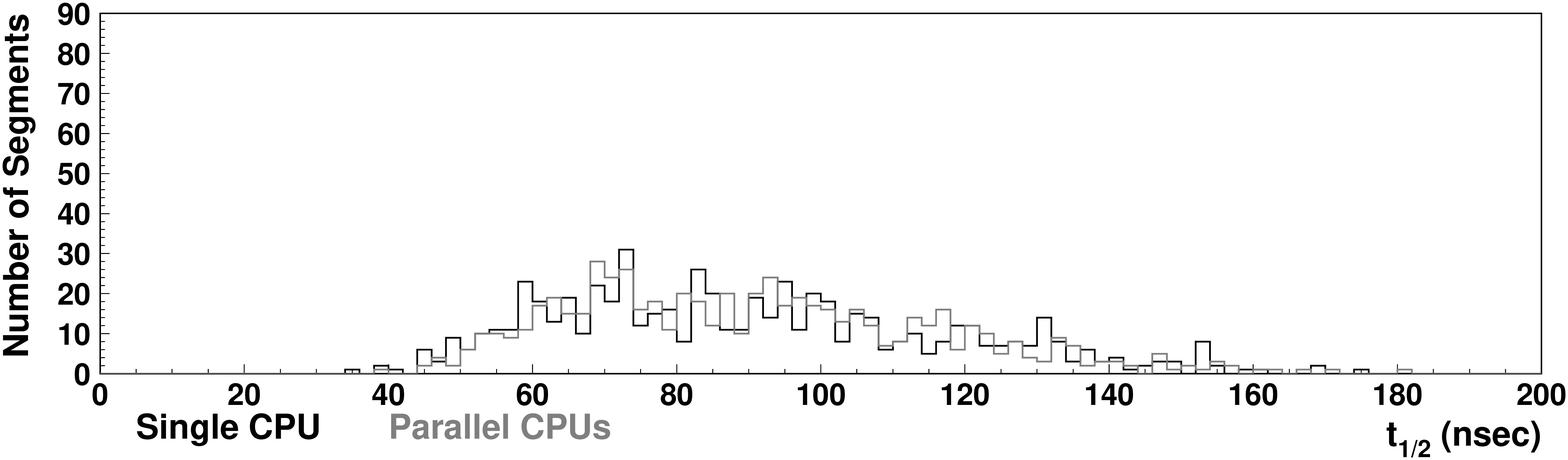}\\
(d)\includegraphics[width=0.96\textwidth]{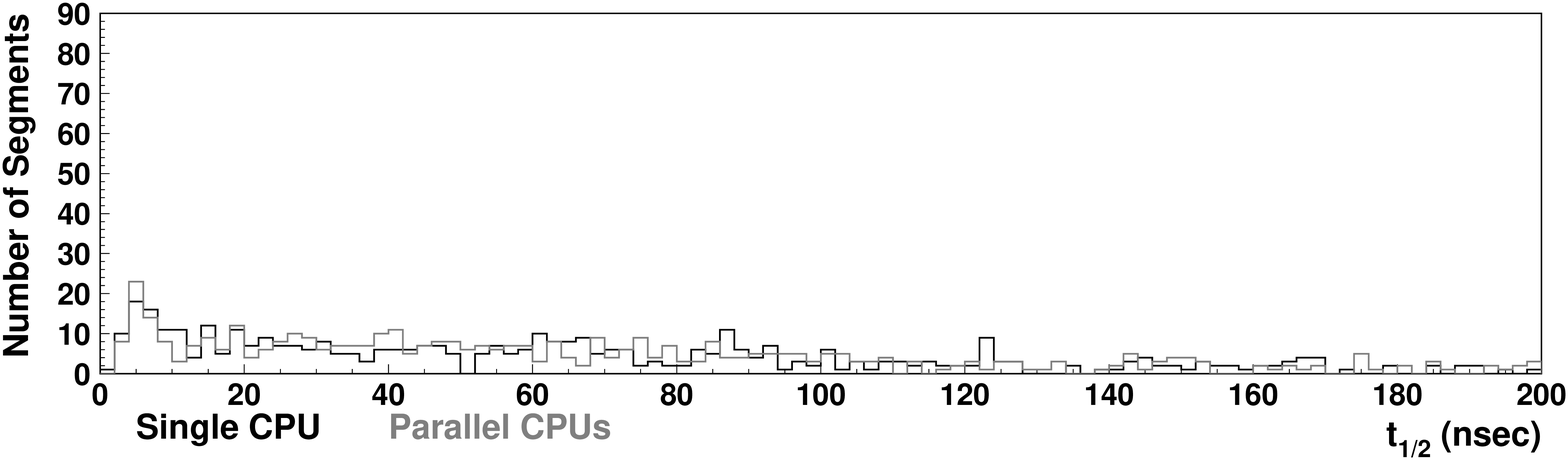}
\end{center}
\caption{Comparison of distribution of median arrival times, $t_{1/2}$, for $2\times 2$~m segments in plane normal to shower trajectory for $10^{16.5}$~eV 
protonic EAS 
simulations with primary zenith angles of a) $0^\circ$, b) $30^\circ$, 
c) $45^\circ$, and d) $60^\circ$.  In each case, $t_{1/2}$ was measured for segments
200~m lateral distance from the shower core. For each 
histogram, good agreement in both mean value and variance are observed 
between simulations generated linearly
(black) and via parallelization (gray). }
\label{fig6}
\end{figure*}

Another comparison that can be performed is to consider secondary particle 
arrival times for specific points at the ground observation level. 
For this purpose,  we consider a ring in the plane normal to the EAS trajectory that intersects the ground at the shower core with   
a radius of 300~m and 2~m thickness. The ring is then further divided 
azimuthally into $\sim1000$ 2-meter long segments.  These segments are then projected onto the ground.  For each segment, we tabulate the arrival time, $t_0$, of the first particle for each segment and the time,
$t_{1/2}$, when 50\% 
of the total particle flux for a given segment has arrived.  These times are
then corrected for the time offset between the positions of each segment on the 
ground and the plane normal to the EAS.  In Figure~\ref{fig5},
we see comparative histograms
of $t_0$ for all four simulation pairs described above.  Figure~\ref{fig6}
shows the same histogram comparisons for $t_{1/2}$.  For both the $t_0$ and 
$t_{1/2}$ comparisons, we see good agreement.  
\section{Conclusion}

Our parallelization technique yields results that are completely 
equivalent to the conventional linear method.  The technique's advantages 
include:
\begin{enumerate}
\item No modification to the underlying simulation routines is necessary 
for this method.  In the case of 
CORSIKA, scripts and binaries were used that were entirely external to 
CORSIKA itself.
\item This technique is highly scalable.  Simulations have been successfully 
executed on systems with more than 3,000 concurrent computational cores.
\item There is also a great deal of flexibility.  Because each job within
a simulation is functionally independent and the duration of each job can be 
specified, it's possible to utilize a wide range of different computational
resources.
\end{enumerate}

There is one major caveat.  While simulations that would have 
previously taken thousands of hours to complete can now be divided into 
thousands of jobs that each take hours to complete, the net use of computational
resources is conserved.  With our current resources, while it is now feasible 
to simulate 100 showers with
primary energies above $10^{19}$~eV, it is not feasible to simulate the tens of 
thousands of showers necessary for a sufficient study of detector response and
acceptance for a UHECR surface array.  Parallelization does, however, provide 
the means to acquire a large reference library of EAS simulations for the 
purpose of developing further techniques.

\section{Acknowledgments}

This work was supported by the U.S.~National Science Foundation awards 
PHY-0601915, PHY-0703893, PHY-0758342, and PHY-0848320 (Utah) 
and PHY-0649681 (Rutgers). 
The simulations presented herein have only been possible due to 
the availability of computational resources at the Center for High Performance 
Computing at the University of Utah which is gratefully acknowledged. A portion
of the 
computational resources for this project has been provided by the U.S.~National 
Institutes of Health (Grant \# NCRR 1 S10 RR17214-01) on the Arches 
Metacluster, administered by the University of Utah Center for High 
Performance Computing.

\bibliographystyle{elsarticle-num}
\bibliography{parallel}

\end{document}